**Running Title:** Long-Distance Flowering Signals

**Corresponding Author:** Ianis G. Matsoukas

School of Life Sciences, Gibbet Hill Campus, The University of Warwick, Coventry CV4 7AL, United Kingdom.

Tel: +44 (0) 24 765 74251, Fax: +44 (0) 24 765 23568

Email: I.Matsoukas@warwick.ac.uk

**Subject Areas:** (1) growth and development, (2) environmental and stress responses

**Number of Black and White Figures:** 0

**Number of Colour Figures:** 1

**Number of Tables:** 1

**Title:** Florigenic And Antiflorigenic Signalling In Plants

**Authors:** Ianis G. Matsoukas, Andrea Juliet Massiah, Brian Thomas

**Authors' addresses:** School of Life Sciences, Gibbet Hill Campus, The University of Warwick, Coventry CV4 7AL. United Kingdom.

**Abbreviations:** ABA, abscisic acid; CC, companion cell; CK, cytokinin; CRY, cryptochrome; FMI, floral meristem identity; FPI, floral pathway integrator; GA, gibberellin; LD, long day; LDP, long day plant; miR, microRNA; Pi, phosphate; PHY, phytochrome; SAM, shoot apical meristem; SD, short day; SDP, short day plant; SE, sieve element; TF, transcription factor; WT, wild type.

**Abstract**

The evidence that FLOWERING LOCUS T (FT) protein and its paralog TWIN SISTER OF FT, act as the long distance floral stimulus, or at least that they are part of it in diverse plant species, has attracted much attention in recent years. Studies to understand the physiological and molecular apparatuses that integrate spatial and temporal signals to regulate developmental transition in plants have occupied countless scientists and have resulted in an unmanageably large amount of research data. Analysis of these data has helped to identify multiple systemic florigenic and antiflorigenic regulators. This study gives an overview of the recent research on gene products, phytohormones and other metabolites that have been demonstrated to have florigenic or antiflorigenic functions in plants.

**Introduction: The Importance of Juvenility and Flowering Time Studies**

Plants undergo a series of qualitative phase transitions during their life cycle in response to environmental and endogenous factors. One of the most distinguishable is the vegetative-to-reproductive phase transition. This stage is preceded by the juvenile-to-adult phase transition within the vegetative phase. During the juvenile phase plants are incompetent to initiate reproductive development and are effectively insensitive to photoperiod or vernalization (acceleration of flowering by a long period of cold temperature). The juvenile phase differs from species to species from a period of a few days, for small herbaceous annual plants, through to periods that may last longer than 20 years (Thomas and Vince-Prue 1997). From a physio-ecological perspective, by having a juvenile phase, plants avoid the low seed yields that would occur if they were to flower precociously while still small and with limited photosynthetic capacity (Thomas and Vince-Prue 1997). With the change to the adult phase of vegetative growth, plants attain competence to respond to floral inducers, which is required for the transition to reproductive phase.

The transition to flowering is a key determinant of plant reproductive success that plays a critical role in the seasonal and geographical adaptation of plants. Understanding the timing of the juvenile-to-adult and vegetative-to-reproductive phase transitions is critical for scheduling in commercial horticulture and arable crops. Recently, emphasis on scheduling flowering has been greatly increased. This is driven primarily by mass marketer demands for product consistency. In addition, increasing cost and price pressures have led scientists to explore molecular and physiological methods to hasten juvenility and time to flowering in order to reduce production costs. Knowledge of how environment influences juvenility and time to flowering could help with crop scheduling in commercial horticulture, decrease time to flowering and reduce waste with resulting benefits for the environment through lower inputs and energy required per unit of marketable product. Furthermore, fast-growing tree species are being increasingly used for pulp and bioenergy production. In such cases, it may be equally important to explore molecular and physiological methods to prevent flowering and prolong juvenility. Moreover, the long juvenile phase length of some species is one of several features limiting efficient breeding programs. Therefore, improving our knowledge of the ways by which species with long juvenile phases regulate their flowering time, it is of great interest, providing approaches to create early flowering phenotypes.

The panoptic theme of physiology and genetics of juvenility and flowering has attracted much attention and many comprehensive review articles have been published (Bernier and Perilleux 2005; Massiah 2007; Turck et al. 2008; Jackson 2009). This review, however, focuses specifically on the multiple molecules that participate in florigenic (floral-promoting) and antiflorigenic signalling in plants. The review is divided into four sections: (i) the first epitomizes aspects of the role of the phloem in long-distance macro- and micromolecular trafficking; (ii) the second illustrates the classical models for explaining the regulation of flowering time; (iii) the third describes the molecular and genetic basis of photoperiodic signalling; and (iv) the fourth details gene products and metabolites that have been demonstrated to have florigenic or antiflorigenic functions in plants.

**The Phloem: A Conduit for the Long-Distance Signalling for Inter-Organ Communication**

Translocation via the phloem provides the most important long-distance transport pathway of the plant. Whilst the xylem tubes transport mainly mineral-containing water from the roots to the shoots, the phloem is responsible for the translocation of organic compounds from the sites of synthesis to the developing and non-photosynthetic tissues. The sieve elements (SEs)-companion cells (CCs) complex is the functional entity responsible for the long-distance phloem translocation not only of photoassimilates, amino acids, organic acids, secondary metabolites, ions, peptides and phytohormones, but also a large range of macromolecules, including proteins, small RNAs and microRNAs (miRs; Turgeon and Wolf 2009; Dinant et al. 2010).

The transport of molecules from the CCs or adjacent parenchyma cells to the SEs takes place either through the apoplasm, based on a series of carriers and pumps, present on the plasma membrane of SEs and CCs, or through the pore-plasmodesma units at the CC-SE interface (Dinant and Lemoine 2010). Most macromolecules present in the SEs are synthesized in the CCs (Turgeon and Wolf 2009). The entry of macromolecules into the SE takes place via the plasmodesmata, which can be either selective or passive via diffusion in a size-dependent manner. The loading of metabolites can follow either symplasmic or apoplasmic routes. Interestingly, it has been proposed (van Bel et al. 2011) that the role of the phloem, including production, release and distribution of signalling molecules, may also encompass modulation and amplification of signals along the long-distance transport conduit.

The driving force for long-distance transport in the SE makes use of a turgor gradient due to variations in photosynthate accumulation along the conduit that create a hydraulic pressure gradient (Knoblauch and Peters 2010). As long distance floral signal transport is now accepted as more complex than the movement of a single type of signal molecule (Bernier and Perilleux 2005), the endogenous compounds translocated within the phloem, along with their involvement in sink-strength regulation may be involved in regulation of the juvenile-to-adult and vegetative-to-reproductive phase transitions.

**Photoperiodism and Models of Flowering Time**

*Photoperiodism in Plants*

For species grown in temperate regions one predictable environmental indicator is the changing duration of daylength. The physiological reaction of plants to the relative length of day and night is known as photoperiodism (Thomas and Vince-Prue 1997). The photoperiodic response allows plant species to adapt to seasonal changes in their local environment. For example, shortening days can prepare for the low temperatures ahead. In 1920, Garner and Allard published that daylength was a major contributor to the time of flowering in a wide variety of plants species (Garner and Allard 1920). Their work laid the foundation for the classification of plant species by their response to photoperiod. Short-day plants (SDPs) flower after the photoperiod becomes shorter than a critical length. Long-day plants (LDPs) show the reverse response, flowering when the scotoperiod becomes shorter than a critical period. Day-neutral plants flower irrespective of the photoperiod/scotoperiod length. Both SDPs and LDPs can be either obligate (also called qualitative) or facultative (quantitative). Obligate plants absolutely require inductive photoperiods to flower, whereas the flowering of facultative plants is only accelerated in the inductive photoperiod; they will still flower under non-inductive photoperiods (Thomas and Vince-Prue 1997). *Arabidopsis thaliana* (hereafter referred as *Arabidopsis*) is a typical facultative LDP. Another photoperiodic response type is represented by intermediate-day plants, which flower if the photoperiod is neither too long nor too short.

In order to distinguish between the short-days (SDs) of spring and autumn, some species have developed a dual daylength requirement. These species require either a series of SDs followed by long-days (LDs) or LDs followed by SDs (Thomas and Vince-Prue 1997). In addition, changing ambient temperatures can alter the daylength requirement of some species. For example *Pharbitis nil* is a SDP in warm conditions but flowers in any daylength at low temperatures (Thomas and Vince-Prue 1997).

*Three Models for Explaining the Regulation of Flowering Time*

Over the years, physiological and biochemical studies have led to three models for explaining the regulation of flowering time. The first is the concept of a universal flowering hormone-like substance, which was first postulated by Mikhail Chailakhyan (reviewed in Thomas and Vince-Prue 1997). The florigen (from Latin, *flora*, "flower"; and Greek, γένεσις, "genesis") concept was based on the transmissibility of floral inductive signals across grafts between reproductive donor stems and juvenile recipients in *Nicotiana tabacum* plants. It was proposed that florigen was synthesized in leaves under inductive daylength conditions and transported to the shoot apical meristem (SAM) via the phloem. The detection of a graft-transmissible floral antagonist also led to the theory of a competing "antiflorigen" (reviewed in Thomas and Vince-Prue 1997). Many years of research were spent researching for the florigen and antiflorigen molecules in the phloem exudates of several plant species, but their molecular character has remained elusive until some recent advances.

The difficulty to dichotomize the hypothetical floral hormone-like substance from phloem-transported assimilates reformed to a second model, the nutrient diversion hypothesis (Thomas and Vince-Prue 1997). This model suggested that floral-promotive conditions result in an increase in the amount of photosynthates translocating to the SAM, which in turn promotes floral induction. The hypothesis that photosynthate translocation is uniquely important in the promotion of floral initiation was displaced by the multifactorial control hypothesis (reviewed in Bernier and Perilleux 2005). This model proposed that several inducers and repressors, including plant hormones and photosynthates, systematize the vegetative-to-reproductive phase transition. According to the multifactorial control hypothesis, floral initiation can only be triggered when the limiting determinants are present at the SAM, at the right dose and time. Furthermore, this model attempted to systematize the diversity of floral responses by suggesting that different genetic, biochemical and physiological factors could be limiting for floral induction in different genotypes and/ or under diverse abiotic and biotic conditions

**Photoperiodic Pathway: Probably the Most Important and Most Conserved of the Floral Induction Pathways**

Flowering time has been genetically explored in several plant model systems, and many loci have been cloned through the study of natural variation and induced mutations. This has led to the conclusion that several interdependent genetic pathways control floral initiation. The photoperiodic and vernalization pathways control time to flowering in response to environmental signals such as daylength, light and temperature, whereas the autonomous and gibberellin (GA)-dependent pathways monitor endogenous indicators of the plant's age and physiological status (Massiah 2007; Turck et al. 2008; Jackson 2009). In addition, other factors and less characterized pathways also play a role in control of floral initiation. These include miRs ( Lee et al. 2010), ethylene (Achard et al. 2007), brassinosteroids (Domagalska et al. 2010), salicylic acid (Wada et al. 2010) and cytokinins (CKs; D'Aloia et al. 2011). Even though described as pathways they cannot be seen as direct networks but rather as parts of a labyrinth since several interactions between the floral pathways have been elucidated. The photoperiodic pathway is of particular interest in this study. Hence, apart from the photoperiodic pathway, these pathways will not be described here. Interestingly, while new aspects of floral induction are continuously uncovered rendering these networks more detailed and interconnected some of the key loci and mechanisms are shared even among distantly related plant species, whereas others are not conserved and give rise to important differences between the plant species.

The actions of all flowering time pathways ultimately converge to control the expression of a small number of so-called floral pathway integrators (FPIs), which include *FLOWERING LOCUS T* (*FT*; Kardailsky et al. 1999; Kobayashi et al. 1999), *TWIN SISTER OF FT* (*TSF*; Yamaguchi et al. 2005), *SUPPRESSOR OF CONSTANS1* (*SOC1*; Yoo et al. 2005) and AGAMOUS-LIKE24 (*AGL24*; Lee et al. 2008; Liu et al. 2008a). These act on floral meristem identity (FMI) genes *LEAFY* (*LFY*; Lee et al. 2008), *FRUITFULL* (*FUL*; Melzer et al. 2008) and *APETALA1* (*AP1*; Abe et al. 2005; Wigge et al. 2005), which result in initiation of flowering.

*Circadian Oscillator Models in Floral Induction*

In order to explain how photoperiod regulates flowering, two models have been proposed: the internal and the external coincidence models (reviewed in Thomas and Vince-Prue 1997). The internal coincidence model proposes that two internal rhythms are in phase under inductive photoperiods and as a result it promotes flowering. Under non-inductive SDs these two rhythms are out of phase and as a consequence flowering is inhibited. On the other hand, the external coincidence model proposes that an external signal (daylength) interacts with an internal light-sensitive rhythm during a certain time of the day. That means that under inductive photoperiods the interaction of light and the light sensitive rhythm occurs and plants are induced to flower. Under non-inductive photoperiods there is no such interaction and flowering is repressed. In addition, floral induction requires not only temporal, but also spatial integration of promotive signal(s).

*Light Perception and Circadian Clock*

The photoperiodic induction pathway (also known as LD promotion pathway) relays light and photoperiodic timing signals to the floral induction process. Light signal has three main functions in the molecular mechanism of the photoperiodic responses: (i) it initiates cues that interact with a circadian oscillator and entrain the circadian rhythm; (ii) it promotes the blue-light-dependent interaction between FLAVIN-BINDING, KELCH REPEAT, F-BOX1 (FKF1) and GIGANTEA (GI); and (iii) it regulates CONSTANS (CO) protein stability.

Several reviews have been published on the circadian clock system recently (Harmer 2009; Imaizumi 2010), so the circadian clock will not be described in great detail here. The circadian clock system has three primary components (Fig. 1). First is the central oscillator/pacemaker that generates the 24h oscillators. A model for the *Arabidopsis* circadian oscillator described a series of multiple interlocked transcriptional–translational feedback loops referred to as the morning, core, and evening loops. Essential components of the core feedback loop are *TIMING OF CAB EXPRESSION1* (*TOC1*), *CIRCADIAN CLOCK ASSOCIATED1* (*CCA1*), and *LATE ELONGATED HYPOCOTYL* (*LHY*; Alabadi et al. 2001; Locke et al. 2005a; Locke et al. 2005b). The morning loop induces *PSEUDO RESPONSE REGULATOR9* (*PRR9*) and *PSEUDO RESPONSE REGULATOR7* (*PRR7*), which are linked to CCA1/LHY (Zeilinger et al. 2006). The evening loop includes GI and ZEITLUPE (ZTL), which are connected to TOC1 in the core loop (Pokhilko et al. 2010). Experimental data and *in silico* analysis (Locke et al. 2005a; Locke et al. 2005b; Pokhilko et al. 2010) have incorporated an unknown component in the core loop, whose activity could be explained by EARLY FLOWERING3 (ELF3; Dixon et al. 2011), EARLY FLOWERING4 (ELF4; McWatters et al. 2007), LUX ARRHYTHMO (LUX; Helfer et al. 2011), and TIME FOR COFFEE (TIC; Hall et al. 2003). The second component is the input pathway that synchronizes or entrains the oscillator with environmental cues. The best-characterized signal is light (reviewed in Kami et al. 2010). In *Arabidopsis,* red/far-red light perception is mediated by PHYTOCHROMES (PHYA, PHYB, PHYD, PHYE). Blue light perception is mediated by CRYPTOCHROMES (CRY1-3; CRY3 binds DNA and is localized in mitochondria and chloroplasts but its role has to be elucidated) and the blue-light sensing proteins ZTL, FKF1 and LOV KELCH PROTEIN2 (LKP2). It has also been suggested that light might act via the photosynthetic chain (Ivleva et al. 2006). Members of each of these photoreceptor families have direct interactions with oscillator genes and proteins. In addition, the plant circadian oscillator is also entrained by daily temperature rhythms but the perception and transduction of such signals is not fully understood (Wenden et al. 2011). The third component is the output pathway that links the oscillator to plant processes under circadian rhythm such as photoperiodic induction.

*The GIGANTEA/ CONSTANS Regulatory Module in Photoperiodic Cascade*

Molecular genetic approaches have identified genes that are responsible for the photoperiodic response. In *Arabidopsis*, transcription of *FT*, which is activated by CO, is a critical aspect of photoperiodic induction (Suarez-Lopez et al. 2001; Wigge et al. 2005; Böhlenius et al. 2006). The CO/FT module is highly conserved in dicot (Jansson and Douglas 2007; Koornneef and Meinke 2010; Srikanth and Schmid 2011) and monocot (Colasanti and Coneva 2009; Distelfeld et al. 2009) plant model systems.

GIGANTEA (GI) has separable roles in regulation of blue light responses, whereas it acts between the circadian oscillator and CO to promote flowering by increasing CO and *FT* mRNA abundance. *Arabidopsis* plants impaired in *GI* locus display reduced transcription levels of *CO* and a late flowering phenotype (Suarez-Lopez et al. 2001). Conversely, expression of *35S::GI* activates transcription of *CO* and *FT* mRNA and confers an early flowering phenotype, even under non-inductive SDs (Mizoguchi et al. 2005). Delicate grafting experiments indicate that CO acts non-cell-autonomously, and can induce flowering when expressed in phloem CCs or in the minor veins of mature leaves, but not in the SAM (An et al. 2004). The *CO* expression pattern is consistent with the external coincidence model, which under LDs activates *FT, TSF* and *SOC1* to promote flowering (Fig. 1; Yamaguchi et al. 2005; Yoo et al. 2005).

Recent advances in understanding CO transcription suggest that CO protein may be stabilized through various mechanisms. A model has been proposed to explain the LD promotion of CO transcription, based on the interaction of GI with the blue-light sensing light protein FKF1, and CYCLING DOF FACTOR1 (CDF1; Sawa et al. 2007). When the FKF1 LOV domain senses blue light in LDs, FKF1 interacts with GI to form a protein complex (Sawa et al. 2007). The GI-FKF1 interaction modulates *CO* mRNA expression, and CO protein stability through degradation of the CO repressor CDF1 (Sawa et al. 2007). CONSTANS (CO) protein is stabilized at the end of the LD period by the activity of PHYA and CRYs. The stability of CO protein is a critical aspect of *FT* induction in LDs (Mizoguchi et al. 2005). Under SD conditions, the diurnal rhythms of *FKF1* and *GI* are out of phase. The lack of interaction between FKF1 and GI with CDF1 in SDs represses flowering via down-regulation of *CO* mRNA (Sawa et al. 2007). Collectively, it seems that GI is not essential for the transcription and diurnal regulation of CO, but rather required to promote CO expression by removing the repression conferred by the CDF proteins. The SUPPRESSOR OF PHYA (SPA) proteins have also been implicated in the stability of CO protein (Laubinger et al. 2006). *Arabidopsis* plants impaired in *SPA* locus flowered early under SDs, mainly due to up-regulation of *FT* expression. Disruption of CO suppresses the early flowering of *spa1* mutant. However, disruption of *SPA1* locus promotes CO protein accumulation, with no detectable effect on *CO* mRNA expression. In addition, it has been shown that the ubiquitin ligase *CONSTITUTIVE PHOTOMORPHOGENIC1* (*COP1*) promotes the degradation of CO protein during the scotoperiod (Jang et al. 2008; Liu et al. 2008b; Lau and Deng 2012). *Arabidopsis* plants impaired in *COP1* show an increase in *FT* expression and an early flowering phenotype under SDs. Impaired function of *CO* partially suppresses the early flowering phenotype of *cop1* mutant. In addition, when grown in the presence of 1% (w/v) sucrose, the *cop1-6* mutant is even able to flower in darkness. In the *cop1* mutant background CO protein is stabilized in the night period, but not in the light period (Jang et al. 2008; Liu et al. 2008b). Collectively, the scotoperiodic-dependent degradation of CO ensures that FT protein is not induced during the scotoperiod (reviewed in Lau and Deng 2012).

Several mechanisms have been demonstrated to have a role in CO-dependent transcriptional activation of FT. The B-box and the CONSTANS, CO-LIKE, TIMING OF CAB 1 (CCT) domains are two of the functional and structural units of CO protein. The CCT domain has been shown to be involved in transcriptional activation of FT by CO. The CCT domain functions as a protein-protein interaction motif through which CO interacts with the HEME ACTIVATOR PROTEIN (HAP) components, HAP3 and HAP5 to comprise a trimeric CCAAT-binding transcription factor complex (Wenkel et al. 2006; Kumimoto et al. 2008; Kumimoto et al. 2010). In addition, CO protein may induce FT transcription by binding to TGACG MOTIF-BINDING FACTOR4 (Song et al. 2008), and to specific *cis*-elements in the FT promoter through its CCT domain (Tiwari et al. 2010). These findings demonstrate the heterogeneous mechanisms by which CO protein may induce FT transcription.

However, expression of *35S::GI* in plants lacking CO and FT can partially rescue their late flowering phenotype, through activation of other pathways independent of CO (Mizoguchi et al. 2005). A mechanism that involves GI regulation of miR172, which mediates photoperiodic flowering in a CO-independent manner has been reported (Jung et al. 2007). In this model, miR172 mediates light signals from GI and promotes flowering in *Arabidopsis* by inducing FT through negatively regulating AP2-like floral repressors, such as TARGET OF EAT1 (TOE1), TOE2, SCHLAFMÜTZE (SMZ) and SCHNARCHZAPFEN (SNZ), but independent of CO (Aukerman and Sakai 2003; Jung et al. 2007). In addition, ectopic expression of *GI* in SDs activates *FT* transcription in a *CO*-independent manner, which involves the FT repressors SHORT VEGETATIVE PHASE (SVP), TEMPRANILLO (TEM) 1, and TEM2 (Sawa and Kay 2011). GIGANTEA was shown to associate with *FT* promoter regions at the same regions bound by SVP, TEM1 and TEM2. GIGANTEA (GI) was also shown to bind to SVP, TEM1 and TEM2 proteins (Sawa and Kay 2011). Thus the mechanisms by which GI promotes *FT* expression independently of CO could include antagonistic binding to the *FT* promoter to prevent FT repressor binding and/or by affecting the stability or activity of the FT repressors, in addition to a mechanism involving miR172.

*The FLOWERING LOCUS T/ TWIN SISTER OF FT Regulatory Module: A Model for the Florigenic Function of FLOWERING LOCUS T*

The FT protein is approximately 23kDa bearing similarity to Raf kinase inhibitory protein of mammals (Chardon and Damerval 2005). *CENTRORADIALIS* (*CEN*), *SELF PRUNING* (*SP*) and *TFL1* are also members of this family, with six members in *Lycopersicon esculentum* and six in *Arabidopsis*. The *Arabidopsis* genome encodes four proteins similar to FT and TFL1, namely TSF, MOTHER OF FT AND TFL1 (MFT), BROTHER OF FT AND TFL1 (BFT) and *Arabidopsis thaliana* RELATIVE OF CENTRORADIALIS (ATC; reviewed in Massiah 2007; Turck et al. 2008; Jackson 2009). FLOWERING LOCUS T (FT) shares 59% of homology with TFL1 (Ahn et al. 2006). Interestingly, although FT antagonizes the activity of TFL1, a modification of a single amino acid can alter TFL1 as an inhibitor of flowering to a promoter of flowering (Hanzawa et al. 2005). In addition, other well characterized floral inhibitors (non-antiflorigens) include FLOWERING LOCUS C (FLC); the canonical output of the vernalization and autonomous pathways, and SVP (reviewed in Massiah 2007; Jackson 2009). Both FLC and SVP form a MADS-box protein complex that represses the transcription of FT through binding to CArG sites in the *FT* gene (Li et al. 2008).

In *Arabidopsis*, expression of FT is regulated in LDs by the photoperiodic pathway (Kardailsky et al. 1999; Kobayashi et al. 1999), the light-quality pathway (Cerdan and Chory 2003; Halliday et al. 2003), the light-quantity pathway (King et al. 2008), the autonomous pathway (Jeong et al. 2009), the GA pathway (Porri et al. 2012), and environmental temperature (Kumar and Wigge 2010; Franklin et al. 2011). High CO expression levels promote *FT* expression in the leaf and its paralog, *TSF*. TWIN SISTER OF FT (TSF) and FT show similar patterns of diurnal oscillation and response to daylengths. However, the spatial expression patterns of TSF and FT do not seem to completely overlap in *Arabidopsis* seedlings, since TSF shown to be expressed in the vascular tissue of hypocotyl and petiole, in the basal part of cotyledons and in the region that will give rise to the SAM. However, as plant development proceeds, FT and TSF expression patterns are more similar (Yamaguchi et al., 2005). Expression of *35S::TSF* confers an early flowering phenotype, similar to *35S::FT* (Mizoguchi et al. 2005). However, *Arabidopsis* plants impaired at the *TSF* locus are not late flowering, but *TSF* disruption enhances the late flowering of the *ft* mutant in the *ft tsf* double mutant background (Michaels et al. 2005; Yamaguchi et al. 2005). Like *FT, TSF* transcription is negatively regulated by FLC (Yamaguchi et al. 2005) and SVP (Jang et al. 2009) and it is repressed by overexpression of SMZ (Mathieu et al. 2009).

Based on the results described above, a model for the photoperiodic induction of *Arabidopsis* has been proposed. In this model, FT protein (Lifschitz and Eshed 2006; Corbesier et al. 2007; Jaeger and Wigge 2007; Lin et al. 2007; Mathieu et al. 2007) and TSF (Michaels et al. 2005; Yamaguchi et al. 2005) are part of the long-distance floral stimulus.

Inductive LD conditions perceived in the leaf stabilize CO protein, which induces *FT* transcription in the leaf (Lifschitz and Eshed 2006; Corbesier et al. 2007), and *TSF* in the stem (Michaels et al. 2005; Yamaguchi et al. 2005). Once translated in the phloem CCs, FT and TSF are loaded to the phloem and translocated to the SAM (Yamaguchi et al. 2005). At the SAM, a series of direct interactions between FPIs, SQUAMOSA PROMOTER BINDING PROTEIN LIKE (SPL) transcription factors (TFs) and FMI genes promote the LD floral induction pathway. FLOWERING LOCUS T (FT) and TSF have been shown to physically interact with the locally transcribed bZIP TF FLOWERING LOCUS D (FD; Abe et al. 2005; Wigge et al. 2005; Yamaguchi et al. 2005; Jang et al. 2009). Phenotypic analysis of double and triple mutants suggest that FT/FD and TSF/FD protein interactions are biologically relevant, as the *fd ft* double mutant flowers later than the *ft* mutant and similarly to the *ft tsf* double mutant and the *ft tsf fd* triple mutant (Jang et al. 2009). The FT/FD transcriptional complex activates the expression of SOC1 (Wigge et al. 2005; Conti and Bradley 2007). Following the induction of *SOC1*, expression of SPL3, SPL4 and SPL5 are rapidly induced in the SAM. These three members of the SPL family are direct targets of SOC1 and FD (Jung et al. 2012), whereas their expression also requires FT/TSF and SOC1/FUL activity (Torti et al. 2012). In turn, the FMI genes AP1, LFY and FUL are directly activated by SPL3 (Wang et al. 2009b; Yamaguchi et al. 2009). Once the transcription of FMI genes is stabilized, FT and TSF are no longer essential and the SAM becomes fully committed to floral initiation (Wigge et al. 2005; Conti and Bradley 2007; Melzer et al. 2008). Under SD conditions, FT expression levels are reduced. However, as plant growth and development proceeds, *FT* expression levels show a clear increase.

Evidence has been provided for the *FT* mRNA trafficking via the phloem to the SAM independently of the FT protein (reviewed in Jackson and Hong 2012). When *FT* mRNA is fused with GFP mRNA or a movement-defective virus, the chimeric virus mRNA moves systemically to the SAM, independent of FT protein (Li et al. 2009; Li et al. 2011a). Whether FT mRNA also participates in systemic floral regulation remains controversial (Li et al. 2009; Li et al. 2011a; Lu et al. 2012). However, under inductive LD conditions, there is another florigenic signal involving specific GAs (Hisamatsu and King 2008; Porri et al. 2012). Thus, there may be multicomponent floral signalling in LDs (Fig. 1) involving FT and GAs, whereas an additional role for photosynthates has also been proposed (Thomas 2006; Hisamatsu and King 2008; King et al. 2008).

*Florigenic and Antiflorigenic Functions of FT Homologous, Orthologous and Paralogous Genes*

Orthologous to *FT* have been identified in several dicot (Böhlenius et al. 2006; Hsu et al. 2006; Lifschitz and Eshed 2006; Lin et al. 2007; Wang et al. 2009a) and monocot (Yan et al. 2006; Komiya et al. 2008) species (Table 1). At least part of the model described above is conserved, with some variations in several plant species. Ectopic over-expression of *FT* orthologous genes hastens the juvenile phase, and promotes time to flowering in transgenic homologous and heterologous plants such as *L. esculentum* (Lifschitz and Eshed 2006)*, Curcubita maxima* (Lin et al. 2007), *Triticum aestivum* (Li and Dubcovsky 2008) and *Oryza sativa* (Komiya et al. 2009). This conservation, together with the small protein size makes FT partially capable of fulfilling the requirements of one of the florigen candidates, or one of its important components.

In *O. sativa,* the function of at least two florigenic molecules: the *HEADING DATE3 ALPHA* (*HD3A*), which promotes flowering in SDs, and *RICE FLOWERING LOCUS T1* (RFT1), which promotes flowering in LDs has been elucidated (Jang et al. 2009; Komiya et al. 2009). Interestingly, FT homologs other than HD3A and RFT1 have been detected in the phloem sap, which suggests the existence of additional florigenic signals in *O. sativa* (Aki et al. 2008). In *Pisum sativum*, evidence suggests the involvement of two *FT* loci in long-distance florigenic signalling. Grafting, expression, and double mutant analyses show that GIGAS/FTa1 regulates a long-distance floral stimulus but also provide evidence for the existence of a second long-distance floral signal that is correlated with expression of FTb2 in leaves (Hecht et al. 2011).

The idea of a certain floral repressor(s) or antiflorigen(s) was proposed almost as early as that of a floral stimulus. Floral repressors are of great importance as they prolong the juvenile-to-adult and vegetative-to-reproductive phase transitions. Their functions allows necessary assimilate reserves to be accumulated, ensuring an unimpeded reproductive development. Interestingly, an antiflorigenic effect of FT-like genes has been postulated in several plant species. *Arabidopsis thaliana* RELATIVE OF CENTRORADIALIS (ATC), an *Arabidopsis* FT paralog function as an antiflorigen (Huang et al. 2012). Functional analysis of the *atc* mutant revealed that ATC acts as a SD induced antiflorigen, which is translocated over long distance and that floral repression by ATC is graft transmissible. *Arabidopsis thaliana* RELATIVE OF CENTRORADIALIS (ATC) probably antagonizes FT activity, because both ATC and FT interact with FD to regulate the same downstream FMI genes, in an opposite manner (Huang et al. 2012). In *Beta vulgaris* time to flowering is regulated by the interplay of two paralogs of *Arabidopsis* FT that have evolved antagonistic functions (Pin et al. 2010). *Beta vulgaris FT2* (*BvFT2*)*,* which is functionally conserved with FT, is essential for floral induction, whereas BvFT1 act as a floral repressor. Similarly, the *Helianthus annuus FT1* (*HaFT1*) paralogue act as a floral promoter, whereas the frame shift allele *HaFT4* act as a floral repressor (Blackman et al. 2010).

*Other Florigenic and Antiflorigenic Regulators*

The *Zea mays INDETERMINATE1* (*ID1*) acts as a floral promoter that is transcribed exclusively in leaf tissue (Colasanti and Sundaresan 2000). *Z. mays* plants impaired at the *ID1* locus have a prolonged vegetative phase. The spatial and temporal expression pattern of *ID1* might indicate its function as a floral inducer that moves from the leaves to the SAM (Colasanti and Sundaresan 2000). Studies on the INDETERMINATE DOMAIN (IDD) TF revealed a connection between the vegetative-to-reproductive phase transition and carbohydrate metabolism-related events (Coneva et al. 2007).

The presence of peptides in phloem exudates is well documented (Stacey et al. 2002). However, the role of peptides and their transporters on flowering time is not well defined. Some early lines of evidence supports the involvement of the plasmalemma-localized *PEPTIDE TRANSPORTER2* (*PTR2*) in regulation of flowering time (Song et al. 1997). Noticeably, *Arabidopsis AtPTR2* antisense plants compared to wild type (WT), exhibit a delay of up to 2 weeks in their flowering time (Song et al. 1997). In addition, in a study for the differential presence of peptides in phloem exudates of flowering and non-flowering *Perilla ocymoides* and *Lupinus albus* plants, several peptides were identified to differ in their abundance (Hoffmann-Benning et al. 2002).

Several studies have identified the antiflorigenic functions of LIKE HETEROCHROMATIN PROTEIN1 (LHP1, TFL2; Gaudin et al. 2001), TEM1 and TEM2 proteins (Castillejo and Pelaz 2008). It has been demonstrated that LHP1, TEM1 and TEM2 function as leaf-based floral repressors that might also be able to move to the SAM. The LIKE HETEROCHROMATIN PROTEIN1 (LHP1) represses the expression of *FT*, but with no effect on the expression of *TSF* (Yamaguchi et al. 2005) or the other FPIs and downstream FMI genes (Gaudin et al. 2001; Takada and Goto 2003; Nakahigashi et al. 2005). TEMPRANILLO1 (TEM1) and TEM2 genes play a key role in inhibiting flowering under SDs and LDs, directly repressing FT and GA biosynthesis genes (Castillejo and Pelaz 2008; Osnato et al. 2012).

**Florigenic and Antiflorigenic Functions of miRNAs**

MicroRNAs (miRs) are short (21-24 nt) non-translated RNAs that are processed by Dicer-like proteins from large, characteristically folded precursor molecules. The majority of plant miRs target TFs and are therefore hypothesised to mainly regulate several developmental processes, such as juvenility and flowering. Most miRs are considered to act in a locally restricted manner, but delicate grafting studies have recently shown that the effect of several miRs is transmissible via grafts (Buhtz et al. 2010) suggesting their transportability.

*The Contrasting Transcriptional Pattern of Phloem-Transmitted miR156 and miR172*

MicroRNA156 (miR156), an ambient temperature-responsive miR (Lee et al. 2010) and strong floral inhibitor, is one of the central regulators of the juvenile-to-adult and vegetative-to-reproductive phase transitions in several plant species (Wu and Poethig 2006; Chuck et al. 2007). Constitutive expression of miR156 prolongs juvenility and time to flowering. Functional analysis of the *hasty1* (*hst1*) mutant of *Arabidopsis* revealed the function of the contrasting transcriptional pattern of the phloem-transmitted miR156 and miR172 (Martin et al. 2009; Varkonyi-Gasic et al. 2010) in regulation of phase transitions (Wu and Poethig 2006; Chuck et al. 2007; Jung et al. 2007; Mathieu et al. 2009). It has been demonstrated that the juvenile-to-adult phase transition is accompanied by a decrease of miR156 abundance, and a concomitant increase in abundance of miR172 and *SPL* TFs. Expression of miR172 activates *FT* transcription in leaves through repression of AP2-like TFs SMZ, SNZ and TOE1-3 (Jung et al. 2007; Mathieu et al. 2009), whereas the increase in *SPLs* at the SAM leads to the transcription of FMI genes (Wang et al. 2009b; Yamaguchi et al. 2009). The FMI genes trigger the expression of floral organ identity genes (Causier et al. 2010), which function in a combinatorial fashion to specify floral organ identities.

Several miR species have been implicated in regulating flowering time in response to carbohydrate metabolism-related events. *Corngrass1* (*CG1*) is a tandem *miR156* locus, which has been identified in several species of the Poaceae family (Schwab et al. 2005; Xie et al. 2006). Overexpression of *miR156/CG1* prolongs juvenility and delays time to flowering (Chuck et al. 2007; Gandikota et al. 2007). Interestingly, overexpression of the *Z. mays miR156/CG1* gene in *Panicum virgatum* confers absence of flowering, even after more than 24 months of growing (Chuck et al. 2011). Since *UBI::CG1* transgenic lines contain greater than two-fold more starch, the lack of flowering in *UBI::miR156/CG1 P. virgatum* overexpression transgenic lines has been associated with starch catabolism-related events (Chuck et al. 2011).

*The Graft-Transmissible miRNA399 Might Regulate Floral Induction in Response to Nutritional Status*

There is a growing body of evidence linking miRs to regulation of nutritional balance in plants and particularly to changes in phosphate (Pi) and sucrose, to floral induction. A recently proposed unique aspect of miR signalling involves potential cross talk between sucrose availability and transport, and miR399 expression under Pi deficiency (Liu et al. 2010). It was shown that when Pi-sufficient plants were subjected to Pi starvation, miR399, an ambient temperature-responsive miR was strongly induced. Expression analysis indicated that expression of TSF was increased in miR399b-overexpressing, and *PHOSPHATE2* (*PHO2*) impaired plants at 23°C, suggesting that their early flowering phenotype is associated with TSF upregulation (Kim et al. 2011). In addition, the induction of *miR399* in either shoots or roots required photosynthetic carbon assimilation. Interestingly, *SUCROSE-PROTON SYMPORTER2* (*SUC2*), a locus encoding a sucrose transporter for phloem loading is inactivated in the *pho3* mutant (Lloyd and Zakhleniuk 2004). The fact that transgenic plants with impaired function of *SUC* genes show alterations in flowering time (Hackel et al. 2006; Sivitz et al. 2007; Chincinska et al. 2008), might suggest that a restriction in photosynthate availability interferes with at least some aspects of Pi deprivation and *miR399* expression levels in regulation of flowering via TSF.

**Florigenic and Antiflorigenic Functions of Phytohormones**
24

Among the phytohormones, GAs are of special importance in their ability to induce flowering in LD plants grown under non-inductive SDs. In *L. temulentum,* $GA_5$ and $GA_6$ have been demonstrated to be LD mobile floral signals that traffic to the SAM (King et al. 2001). In *Arabidopsis*, the highly bioactive $GA_1$ and $GA_4$ show strong florigenic activities. The *GIBBERELLIC ACID-INSENSITIVE* (*GAI*) mRNA, a critical component of GA signal transduction in *Arabidopsis* has been shown to undergo long-distance transport (Haywood et al. 2005). Certain 2-oxidase-resistant isoforms of GA also regulate floral initiation by their transport to the SAM. In a GA20-oxidase2 silenced line of *Arabidopsis* reduced GA biosynthesis delayed flowering but only in far-red LD conditions (Hisamatsu and King 2008). Furthermore, GAs have important roles in promoting transcription of *FT, TSF* and *SPL* genes during floral induction in response to LDs (Porri et al. 2012). These functions are spatially separated between the leaf and SAM (Porri et al. 2012).

Interestingly, a mechanistic basis for the interaction between the photoperiodic, autonomous and GA pathways is suggested by the convergence of the three pathways on the promotion of FMI genes (Blazquez and Weigel 2000; Achard et al. 2006; Eriksson et al. 2006). Gibberellin$_4$ ($GA_4$) promotes export of assimilates, and in combination with sucrose has a synergistic effect on the activation of FMI genes (Blazquez and Weigel 2000; Eriksson et al. 2006). The increase of $GA_4$ and sucrose at the SAM was shown to be due to their transport from exogenous sources to the SAM via the phloem (Eriksson et al. 2006).

Cytokinins (CKs), is another class of phytohormones with reported roles in modulation of flowering time in *Arabidopsis* and *S. alba*. Mutation at the *ALTERED MERISTEM PROGRAM1-1* locus of *Arabidopsis* increases CK content and confers, amongst many other phenotypes, early flowering (Griffiths et al. 2011). Phloem sap analyses in *Arabidopsis* have revealed increased levels of isopentenyl type of CKs during floral initiation, in response to photoperiod (Corbesier et al. 2003). Supply of CK in the form of benzylaminopurine for 8 h in the root system of *Arabidopsis* plants grown in hydroponic culture, was sufficient to induce flowering in non-inductive SDs. The florigenic effect of CK, which bypasses *FT*, acts via its paralogue *TSF,* and *SOC1* (D'Aloia et al. 2011). When CKs are applied to the SAM of the LDP *S. alba*, they confer an increase in mitotic activity similar to that conferred by exposure to inductive LDs, and *SOC1* up-regulation (Bernier et al. 1993; Bernier 2011). Cytokinins (CKs) and *SaFT* may therefore be an integral part of the floral stimulus in *S. alba*. It is possible that CKs and *SaFT* act synergistically at the SAM; the CK being responsible for *SaSOC1* up-regulation and mitotic activation, and *SaFT* triggering downstream regulators of *SaSOC1* (Bernier 2011).

*Other Potential Long-Distance Phytohormonal Signals on Flowering*

Abscisic acid (ABA) was proposed as the first identified antiflorigen substance, and hence is regarded as a floral repressor (reviewed in Thomas and Vince-Prue 1997). *Arabidopsis* mutants defective in or insensitive to ABA flower early, whereas mutants overproduce ABA flower late (Achard et al. 2006; Domagalska et al. 2010; Quiroz-Figueroa et al. 2010). The inhibitory effect of ABA on time to flowering might be explained via sugar repression-related events. This is supported by the early flowering phenotype of ABA deficient mutants and their allelism to sugar-insensitive mutants (Rolland et al. 2006). Despite its floral inhibitory effects in some plant species, however, many others are not affected, and therefore ABA seems to have no general function as a floral transmissible repressor.

Polyamines such as putrescine and spermidine might also represent part of the long-sought florigen. Photoperiodic induction of *S. alba* was correlated with a significant increase in the abundance of putrescine, the major polyamine in the leaf phloem sap (Havelange et al. 1996). Leaves of *S. alba* treated with an inhibitor of putrescine anabolism reduced the levels of putrescine in the phloem sap and also repressed the transition to flowering. This could indicate that putrescine may be an integral part of the floral signal in *S. alba* (Havelange et al. 1996). In addition, polyamines in the form of spermidine might also be involved in SD floral induction of *Arabidopsis*, as a connection between spermidine and time to flowering has been proposed (Applewhite et al. 2000).

Auxins have also been detected in phloem sap (Petrasek and Friml 2009), and in conjunction with other molecules may have potential florigenic activities. Disruptions in any aspects of auxin metabolism-related events including biosynthesis, polar transport and signalling confer dramatic defects in flowering time. Mutation in the *Arabidopsis AUXIN RESPONSE FACTOR2* (*ARF2*), a key locus responsible for auxin-mediated signalling and gene transcription, confers late flowering (Okushima et al. 2005). *Arabidopsis* plants impaired at the *HYPONASTIC LEAVES1* (*HYL1*) locus display delayed time to flowering by altering sensitivity to auxin, CKs and ABA (Lu and Fedoroff 2000). Impaired function of *ACC-RELATED LONG HYPOCOTYL1* (*ALH1*) locus confers late flowering by affecting sensitivity to auxin and ethylene (Vandenbussche et al. 2003). Furthermore, a connection between auxins and GAs in regulation of flowering time has also been proposed. The GA mutant *nana1* (*na1*) of *Pisum sativus,* flower late due to a defect in auxin regulation of GA biosynthesis (DeMason 2005). The altered flowering phenotypes of plants impaired in auxin biosynthesis and signalling might indicate that modulation of flowering time by auxins is indirect, via interaction with other phytohormones, such as CKs, ethylene, ABA and GAs.

Phytohormones such as salicylic acid (Wada et al. 2010), ethylene (Achard et al. 2007) and ascorbic acid (Kotchoni et al. 2009) have also been suggested to act as floral signals. Brassinosteroids have been suggested to be involved in the autonomous pathway, but also to co-operate with GAs in controlling flowering time (Domagalska et al. 2010).

**Sucrose as a Florigen**

Sucrose is the most extensively studied compound that might participate in long-distance signalling for flowering. It is the dominant transport metabolite for long distance carbon transport between source and utilization sinks. Sucrose is synthesized in the cytosol from photosynthetically fixed carbon, starch or lipids. This biochemical process is enzymatically carried out by sucrose-phosphate synthases and sucrose-phosphate phosphatases. In *Arabidopsis* and *S. alba* exposed to inductive LDs sucrose levels increase rapidly and transiently in phloem leaf exudates (Bernier et al. 1993; Corbesier et al. 1998). Increased sucrose export in these studies was shown to derive from starch catabolism-related events. Three independent mutant alleles of *Arabidopsis thaliana SUCROSE TRANSPORTER9* (*AtSUC9*) have been shown to flower earlier than WT under non-inductive SDs. This flowering phenotype is opposite to what is observed when *AtSUC* gene activities are impaired in the phloem. Based on the high affinity of *AtSUC9* for sucrose and general expression patterns and phenotype of *atsuc9* mutants, it is possible that *AtSUC9* prevents early flowering by maintaining a low concentration of extracellular sucrose (Sivitz et al. 2007) indicating a potential link between *AtSUC9* transcription, a particular carbohydrate threshold level and transition to flowering. Down-regulation or over-expression of *SUCROSE TRANSPORTER4* (*AtSUT4*) in *Solanum tuberosum* delays or promotes time to flowering, respectively (Chincinska et al. 2008). Besides flowering time, in the same work evidence was provided on *SUT4* involvement in the shade avoidance response. This suggest that *PHY*-dependent and photoperiod-dependent developmental responses, such as floral initiation and shade avoidance share a common downstream mechanism in which sucrose accumulation levels are actively involved. Defoliation experiments demonstrated that the sucrose export-increase coincides with the start of mobile signal transport, and occurs before the activation of cell division at the SAM (Corbesier et al. 1998). Exogenous sucrose application to *Arabidopsis* WT plants grown under suboptimal photosynthetic conditions promotes floral initiation (Bagnall and King 2001). Time to flowering in *Fuchsia hybrida* is hastened by high light integral levels, even under non-inductive SD conditions. It has been shown that this involves transport and accumulation of sucrose at the SAM (King and Ben-Tal 2001). Flowering in the *uniflora* (*uf*) mutant of *S. lycopersicum* has also been shown to be irradiance-dependent (Dielen et al. 2004). The dramatic late-flowering phenotype of the *uf* mutant under low irradiance suggests that sucrose transport and accumulation at the SAM might be an essential signal for floral initiation in *S. lycopersicum*.

One possible role for sucrose could be to act on *FT* with direct effects on its expression. Sucrose can be supplied to the aerial part of plants by growing them on vertical plates. Under these conditions, sucrose almost completely suppresses the late-flowering phenotype of mutants, such as *gi3, co3* and *fca,* but is unable to rescue *ft* and *fwa* (Roldan et al. 1999). This suggests a sucrose interaction, upstream of *FT*. Photosynthetic promotion of floral initiation under LDs via *FT* has also been demonstrated by King *et al.,* (2008). Additional support on photosynthate involvement upstream of *FT* is provided by the interaction of several photoperiodic mutants with elevated [$CO_2$] (Song et al. 2009). Recently, the *IDD8* locus of *Arabidopsis* was reported to have a role in *FT*-dependent induction of flowering by modulating sugar transport and metabolism via the regulation of *SUCROSE SYNTHASE4* activity (Seo et al. 2011). In SDs the late-flowering phenotype of starch-deficient mutants, can be rescued by low levels of exogenous sucrose application (Corbesier et al. 1998; Yu et al. 2000; Xiong et al. 2009). Under LD conditions, Ohto *et al.,* (2001) observed an 8 d delay in flowering of *Arabidopsis* plants grown on medium containing 5% compared with 1% (w/v) sucrose. This response coincided with a significant increase in the number of rosette leaves and plant size at flowering. Furthermore, the time-period that plants received more sucrose increased the magnitude of the delay, suggesting a threshold level for the response to flowering to additional sucrose (Ohto et al. 2001). However, elevated sucrose levels can slightly delay flowering and reduce *FT* and *SOC1* mRNA transcription levels. In the *co* mutant background, sucrose accelerates flowering by-passing *FT* and *SOC1* (Ohto et al. 2001). Hastening of the time to flowering might be via *LFY* activation by a synergistic interaction between sucrose and GA at the SAM (Blazquez and Weigel 2000; Eriksson et al. 2006).

Sucrose affects skotomorphogenic time to flowering. Several *Arabidopsis* mutants such as *gi*, *co*, *luminidependens* (*ld*) and *cop1* can overcome skotomorphogenesis and flower if sucrose is supplied (Roldan et al. 1999; Nakagawa and Komeda 2004). In this case, sucrose regulates time to flowering by acting as an energy source. However, the promotive effect of sucrose under skotomorphogenic conditions may also be concentration-dependent. *Arabidopsis* plants grown in the dark on medium containing 6% (w/v) sucrose exhibit a late flowing phenotype, compared with plants grown on 2% (w/v) sucrose (Zhou et al. 1998).

Trehalose-6-phosphate (Tre6P) is a metabolite of emerging significance with hormone-like metabolic activities. The level of Tre6P is closely linked to the level of sucrose, which supports the hypothesis that Tre6P acts as a signal of sucrose status (Lunn et al. 2006). Trehalose-6-phosphate (Tre6P) is formed during the synthesis of trehalose by trehalose-phosphate synthase (TPS) and trehalose-phosphatase (TPP; Rolland et al. 2006). The association of Tre6P activity with floral induction is demonstrated by the function of the *TPS1* gene. Mutation of *TPS1* by insertional mutagenesis in *Arabidopsis* disturbs, amongst many other unprecedented phenotypes, the vegetative-to-reproductive phase transition (Dijken et al. 2004). Interestingly, *tps1* mutant plants completely fail to flower, unless TPS1 is induced by means of dexamethasone-inducible *TPS1* expression. In addition, evidence indicates that Tre6P acts as an intermediate between sucrose and ADP-glucose pyrophosphorylase (Kolbe et al. 2005), the key enzyme of starch synthesis in leaves, and that the Tre6P signal is integrated into the miR156/SPL node of the floral induction pathway (Markus Schmid and Mark Stitt, unpublished data). Given the evidence that Tre6P acts as a sucrose signal to regulate the vegetative-to-reproductive phase transition, this might provide a route whereby starch metabolism-related events are linked to the sucrose status in leaves, and transition to flowering.

However, studies based on developmental transitions in response to sucrose metabolism-related events are still in their infancy. This is related to the complexity of the source-sink interactions and the wide-range effect of sucrose, which is clouded by its function as nutrient, signalling molecule, osmotic regulator, and by the interaction between sucrose signalling, hormone and nutrients networks.

**Conclusions and Perspectives**

The juvenile-to-adult and vegetative-to-reproductive phase transitions are key determinants of plant reproductive success. The most striking advances in our understanding of the genetic regulation of plant developmental transitions have derived from studies of flowering-time in several dicot and monocot plant model systems. The transition to flowering involves multiple factors and pathways, which show different responses to environmental and endogenous factors. Much of the evidence for both floral-inducing substances and floral inhibitors can be subject to multiple interpretations. Nevertheless, it can be proposed that the prolonged juvenile-to-adult and vegetative-to-reproductive phase transitions might be due to a plethora of antiflorigenic signals such as miR156, LHP1, ATC, and TEM proteins. In any case, the transcription levels of *FT* and its homologues can account for many of the reported results. The ability of FT protein to act as a graft-transmissible floral signal in diverse plant species has significant scientific, economic and practical implications in plant science. Although the FT gene and its transcription products are somewhat understood, many issues still need to be further characterized. For instance, the difference in the functions and spatial and temporal expression of the *FT/TSF* protein complex between gymnosperms and angiosperms species, and between annual and perennial plant species remains to be further characterized. Furthermore, long distance floral signal transport is now accepted as more complex than the movement of a single type of signal molecule. Hence, the effect of other endogenous compounds translocated within the phloem in regulation of the juvenile-to-adult and vegetative-to-reproductive phase transitions, is another issue to be elucidated.

**Acknowledgements**

The authors would like to acknowledge support from the Hellenic State Scholarships Foundation (IKY) and UK Department for Environment, Food and Rural Affairs (DEFRA) in preparing this review.

Figure 1: Florigenic and antiflorigenic signalling pathways in *Arabidopsis*. Components of the signalling pathways are grouped into those that promote (↓) and those that enable (⊤) the transition to flowering. The main florigenic and antiflorigenic components and interactions are depicted in the diagram, but additional elements have been omitted for simplicity. Details are given in the text.

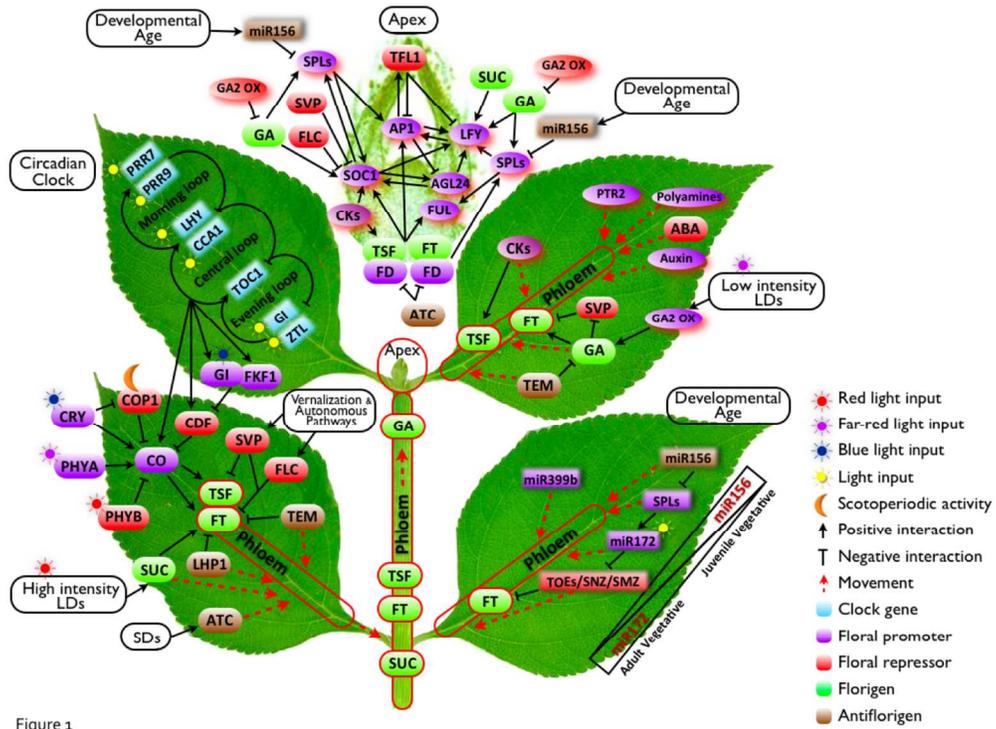

Figure 1

508x381mm (72 x 72 DPI)

Table 1: List of *FLOWERING LOCUS T* loci in annual, biennial and perennial plant species.

| Classification | FT loci | Mutant phenotype | Overexpression phenotype | Expression in *Arabidopsis* | Reference |
|---|---|---|---|---|---|
| **Annuals** | | | | | |
| *Arabidopsis thaliana* | *FT* | Late flowering | Early flowering | _ | (Kardailsky et al. 1999) |
| | *TSF* | No phenotype | Early flowering | _ | (Kobayashi et al. 1999) |
| *Cucurbita maxima* | *CmFTL1* | _ | _ | Early flowering | |
| | *CmFTL2* | _ | _ | Early flowering | (Lin et al. 2007) |
| *Gentiana capitata* | *GtFT1* | _ | Early flowering | Early flowering | |
| | *FtFT2* | _ | Early flowering | Early flowering | (Imamura et al. 2011) |
| *Helianthus annuus* | *HaFT1* (Wild type) | _ | _ | Late flowering | |
| | *HaFT1* (Cultivated) | _ | _ | Early flowering | (Blackman et al. 2010) |
| | *HaFT2* | _ | _ | Early flowering | |
| | *HaFT4* | _ | _ | Early flowering | |
| *Pharbitis nil* | *PnFT1* | _ | Early flowering | Early flowering | |
| | *PnFT2* | _ | _ | _ | (Hayama et al. 2007) |
| *Lactuca sativa* | *Ls*FT | _ | _ | Early flowering | (Fukuda et al. 2011) |

| Medicago truncatula | MtFTa1 | Late flowering | Early flowering | Early flowering | |
|---|---|---|---|---|---|
| | MtFTa2 | – | – | No phenotype | |
| | MtFTb1 | – | – | Early flowering | (Laurie et al. 2011) |
| | MtFTb1 | – | – | No phenotype | |
| | MtFTc | No phenotype | – | Early flowering | |
| Oryza sativa | HD3A | Late flowering in SDs | Early flowering | Early flowering | (Komiya et al. 2009; Li et al. 2011b) |
| | RFT1 | Late flowering in LDs | Early flowering | Early flowering | (Jang et al. 2009; Komiya et al. 2009) |
| Pisum sativum | PsFTa1/ GIGAS | Late flowering | – | Early flowering | |
| | PsFTa2 | – | – | Early flowering | |
| | PsFTb1 | – | – | Early flowering | (Hecht et al. 2011) |
| | PsFTb2 | – | – | Early flowering | |
| | PsFTc | – | – | Early flowering | |
| Solanum lycopersicum | SP3D/SFT | Late flowering | Early flowering | – | |
| | SP5G | – | – | – | (Lifschitz and Eshed 2006; Shalit et al. 2009; Krieger et al. 2010) |
| | SP6A (Wild type) | No phenotype | – | – | |
| *Triticum aestivum* | *TaFT/VRN3* | Late flowering | Early flowering | _ | (Yan et al. 2006) |
| *Zea mays* | *ZCN8* | Late flowering | Early flowering | Early flowering | (Danilevskaya et al. 2011; Lazakis et al. 2011; Meng et al. 2011) |
| **Biennials** | | | | | |
| *Beta vulgaris* | *BFT1* | No flowering | Late flowering | _ | |
| | *BFT2* | Late flowering | Early flowering | Early flowering | (Pin et al. 2010) |
| **Perennials** | | | | | |
| *Citrus × sinensis* | *CiFT* | _ | _ | Early flowering | (Endo et al. 2005) |
| *Malus domestica* | *MdFT1* | _ | Early flowering | Early flowering | (Kotoda et al. 2010; Trankner et al. 2010) |
| | *MdFT2* | _ | _ | _ | (Kotoda et al. 2010) |
| Oncidium 'Gower Ramsey' | *OnFT* | _ | _ | Early flowering | (Hou and Yang 2009) |
| *Populus tremula* | *PtFT1* | _ | Early flowering | Early flowering | (Böhlenius et al. 2006; Hsu et al. 2006; Hsu et al. 2011) |
| | *PtFT2* | _ | Early flowering | Early flowering | |
| *Solanum tuberosum* | *StSP3D* | Late flowering | _ | Early flowering | |
| | *StSP5G* | _ | _ | _ | (Navarro et al. 2011) |
| | *StSP6A* | No phenotype | Early flowering | _ | |
| *Vitis vinifera* | *VvFT* | – | – | Early flowering | (Sreekantan and Thomas 2006; Carmona et al. 2007) |
| *Chrysanthemum seticuspe* | *CsFTL1* | – | Early flowering | – | |
| | *CsFTL2* | – | No phenotype | – | (Oda et al. 2012) |
| | *CsFTL3* | – | No phenotype | – | |

4